# Observation of Rayleigh scattering by simplified optical correlation-domain reflectometry without frequency shifter

Tomoya Miyamae[1], Zhu Guangtao[1], Takaki Kiyozumi[1], Kohei Noda[1,2], Heeyoung Lee[3], Kentaro Nakamura[2], and Yosuke Mizuno[1*]

[1] *Faculty of Engineering, Yokohama National University, Yokohama 240-8501, Japan*

[2] *Institute of Innovative Research, Tokyo Institute of Technology, Yokohama 226-8503, Japan*

[3] *Graduate School of Engineering and Science, Shibaura Institute of Technology, Tokyo 135-8548, Japan*

[*]E-mail: mizuno-yosuke-rg@ynu.ac.jp

We present a method for measuring the transmission loss of an optical fiber using simplified optical correlation-domain reflectometry. By intentionally lowering the spatial resolution of the system, we observe the Rayleigh scattered signal for the first time without the need for a frequency shifter. Using this method, we simultaneously measure the transmission loss, location, and loss of faulty connections along a 10-km-long fiber under test by performing distributed reflected-power measurements of Rayleigh scattering and Fresnel reflection.

In recent years, the rapid growth of fiber-optic communication has led to increased demand for health monitoring techniques for fiber networks [1]. Optical reflectometry detects faults and damages in the fiber by measuring the reflectivity distribution along a fiber under test (FUT). It has various advantages over the traditional electrical sensors, such as compact size, long measurement range, distributed measurement, strong resistance to electromagnetic interference, and the absence of a requirement for powering the sensing component.

Optical reflectometers can be classified based on their operating principles: optical time-domain reflectometry (OTDR) [2-7] and optical frequency-domain reflectometry (OFDR) [8-14] are widely known for their long measurement range and high spatial resolution, respectively. However, OTDR requires the repetition of thousands of pulse injections and data processing to achieve a high signal-to-noise ratio (SNR), resulting in a relatively low sampling rate. On the other hand, OFDR requires a high-quality laser due to the trade-off between the laser linewidth and the measurement range. Although some improvements have been reported [12], an additional interferometer is required, resulting in a complex and expensive experimental setup.

As an alternative to OTDR and OFDR, optical correlation-domain reflectometry (OCDR) [15-33] has been developed on the basis of the synthesis of optical coherence functions (SOCF) [15]. OCDR can be implemented at a relatively low cost and has a unique advantage called random accessibility, i.e., the ability to access an arbitrary sensing position on the FUT at extremely high speed. In standard OCDR, an acousto-optic modulator (AOM) is used to shift the frequency of the reflected (or reference) light from the FUT by several tens of megahertz for the suppression of low-frequency noise during the heterodyne detection [20]. To date, several approaches have been developed to simplify the standard OCDR for the reduction in implementation cost. One is an AOM-free configuration, which operates by monitoring the foot of the reflected peak in the spectrum [26]. Other types of simplified configurations without the direct use of a reference path have also been reported [27–29]. To further simplify the OCDR system, we have recently developed a new configuration that eliminates the need for an electrical spectrum analyzer (ESA), which builds upon the AOM-free OCDR system and boasts enhanced spatial resolution and operational speed [30]. However, these improvements come at a cost of significant noise for long-range measurements, as has been reported previously [31]. Here, we focus on the AOM-free OCDR system with the use of an ESA.

In general, by detecting the Rayleigh scattered signal returned from the FUT in distributed reflectivity sensors, the light transmission loss and the position of the fiber end can be determined. In previous research, successful observation of Rayleigh scattering in a standard

OCDR has been reported, while no estimation of the transmission loss has been done. In an AOM-free OCDR, there has been no reports concerning either due to not only the optimization of experimental conditions for clear detection of Fresnel reflection peaks that eliminates unrelated signals, but also the tendency for Rayleigh signals to be buried by low-frequency noise. Therefore, it is essential to employ special measures to enable the observation of Rayleigh scattering in the AOM-free OCDR.

In this work, we demonstrate the observation of Rayleigh scattering signals in an AOM-free OCDR by intentionally deteriorating the spatial resolution. First, we experimentally confirm the inversely proportional relationship between the modulation amplitude and the spatial resolution during the distributed measurement along a 1-km-lomg FUT. Subsequently, the dependence of the sensitivity of Rayleigh scattering observation on varied spatial resolution values is investigated, verifying the effectiveness of our approach. Distributed measurement along a 10-km-long FUT is then conducted and the transmission loss is determined by analyzing the Rayleigh signal. Finally, we combine simultaneous detection of both Fresnel reflection and Rayleigh scattering, verifying the capability of AOM-free OCDR to operate in a similar manner to other distributed reflectivity sensors.

The standard experimental setup for OCDR is shown in Fig. 1. The output light from a laser is divided into two light beams, with one the incident to the FUT and the other the reference used as a local oscillator. The interference of the signal and reference is then detected with a photodetector (PD), and the electrical beat signal is observed in an ESA. The zero-span function of the ESA is used to track the temporal change of the light power of a fixed frequency at the foot of the spectrum, which is finally displayed on an oscilloscope (OSC).

To perform distributed measurement along the FUT, the frequency of the laser output is modulated in a sinusoidal waveform by directly controlling the driving current, which can be expressed as [18-21]

$$f(t) = f_0 + \Delta f sin(2\pi f_m t), \tag{1}$$

where $f_0$ is the center frequency, $\Delta f$ is the modulation amplitude, and $f_m$ is the modulation frequency. As a result, the coherence function [19] of the interfering lights is synthesized to a delta-function-like correlation peak within the FUT [23], which is used to selectively observe the reflected light at the sensing position. By sweeping the modulation frequency, the correlation peak can be scanned along the FUT, and thus the reflectivity can be measured in a distributed manner.

Figures 2(a) and 2(b) depict the observed reflection spectra in the standard OCDR and AOM-free OCDR, respectively. During the heterodyne detection in a standard OCDR, the optical frequency of the reference light (or reflected light) is shifted by several tens of megahertz to avoid the low-frequency disturbance caused by the PD and the ESA, as shown in Fig. 2(a). In the AOM-free OCDR, the reflection peak appears in the 0-Hz band with a finite linewidth which is dependent on the laser linewidth. By observing the foot of the reflection spectrum, the overlapping between the reflected signal and the low-frequency disturbance can be avoided, as shown in Fig. 2(b). Note that the measurement range $D$ and the spatial resolution $\Delta z$ are given in the same manner as those in standard OCDR by Eqs. (2) and (3) [33]:

$$D = \frac{c}{2nf_m}, \tag{2}$$

$$\Delta z \cong \frac{2f_z c}{\pi n f_m \Delta f}, \tag{3}$$

where $c$ is the light velocity in vacuum, $n$ is the refractive index of the fiber core, and $f_z$ is the center frequency of the zero-span function of the ESA. The observation of Rayleigh scattered signals has proven to be challenging in AOM-free OCDR due to the tendency of the weak signal to be buried by low-frequency disturbance. To improve the sensitivity to the Rayleigh scattering, we intentionally deteriorate the spatial resolution of the system by decreasing the modulation amplitude of the laser driving current, controlled by a function generator (FG). This approach is based on the inversely proportional relationship between spatial resolution and modulation amplitude, as expressed in Eq. (3).

The experimental setup of the AOM-free OCDR implemented for the distributed observation of Rayleigh scattering is shown in Fig. 3. The length of the delay line was 6 km. The modulation frequency was scanned from 32.5 kHz to 58.5 kHz to perform distributed measurement along a 1-km-long FUT. The sweeping period was 10 ms (repetition rate: 100 Hz). The wavelength of the laser output was 1550 nm with a ~2-MHz linewidth. To better observe the reflected Rayleigh scattered signal, some of the experimental conditions were set differently from those of the conventional OCDR: the center frequency of the zero-span function of the ESA was set to a relatively large value of 1.4 MHz; the resolution bandwidth was 300 kHz, and the video bandwidth was off. In addition, the built-in band-pass filter of the PD was set to a wide band of 0-10 MHz. Note that even under these experimental conditions, it was still difficult to observe the Rayleigh signal unless the spatial resolution was intentionally deteriorated.

First, we experimentally verified the inversely proportional relationship between the spatial resolution and the modulation amplitude. The structure of the FUT used in the experiment is shown in Fig. 4(a). Distributed measurement of the reflected spectrum from the end of the FUT was performed under the conditions of successively changed modulation frequency from 1.6 GHz to 19.3 GHz controlled by the FG. The Lorentzian fitting results of the reflected spectrum are shown in Fig. 4(b). The horizontal axis shows the relative position along the FUT, whose origin is the position of the angled-physical-contact (APC) connector between the circulator and the FUT. The vertical axis was normalized by setting the average power of the noise floor 0 and the peak power 1. From the purple curve to the red, the modulation amplitudes were 1.6, 2.2, 3.9, 7.5, and 19.3 GHz, respectively. Figure 4(c) shows the measured spatial resolution values under varied modulation amplitudes. It is verified that as the modulation amplitude decreases, the full width at half-maximum (FWHM) of the reflection peak, i.e. spatial resolution of the system, is inverse-proportionally deteriorated.

The results of distributed reflectivity measurement along the entire FUT under varied spatial resolution values are shown in Fig. 4(d). From the red curve to the purple, the spatial resolution values are 0.26, 0.38, 0.49, 0.83, and 1.48 m, respectively. While successfully detecting the end of the FUT at ~ 1 km, the results also show that the Rayleigh scattering can be more easily observed under deteriorated spatial resolution values. With the relative power defined as the subtraction of the average power of the Rayleigh signal to the noise floor, its values under varied measured spatial resolution values are shown in Fig. 4(e). It is shown that the relative power of the Rayleigh signal increases as the spatial resolution deteriorates because of the enlarged detection range. However, as the loss of standard silica single-mode fibers can be as low as 0.2 dB/km at 1550 nm, the power attenuation of the Rayleigh scattering signal is not clearly observed in this experiment, as the measurement range is relatively short.

To estimate the transmission loss of the fiber, we then increased the measurement range to approximately 10 km. The structure of the FUT is shown in Fig. 5(a). The modulation frequency was swept from 4.26 kHz to 8.50 kHz, and the modulation amplitude was 1.6 GHz. The center frequency of the zero-span function of the ESA was 2 MHz. The length of the delay line in the reference path was 50 km. The result of the distributed measurement is shown in Fig. 5(b). With longer measurement range, clear attenuation of the Rayleigh scattering signal is observed along the transmission direction. The attenuation appeared in the figure is linearly approximated (red line), and half of its slope gives the transmission loss of the fiber of 0.181 dB/km. The measured value is in good agreement with the nominal value of the FUT of 0.186 dB/km.

Finally, distributed reflectivity measurement was performed using an FUT consisting of multiple long fibers. The structure of the FUT is shown in Fig. 6(a), where fiber segments of. 0.5 m, 4.6 km, 0.5 m, 3.0 km, and 2.0 km are connected in sequence. At the ends of the 4.6-km fiber, physical-contact (PC) connectors are intentionally used, while APC connectors are used at other joints. Note that the reflectivity at a PC connector is much higher than that at an APC connector. The modulation frequency was swept from 4.26 kHz to 8.35 kHz. Loss was intentionally applied by loosening the APC connection before the 3.0 km fiber to simulate a fault in a fiber communication network. The result of the distributed measurement is shown in Fig. 6(b). From the dark blue curve to the light blue curve, the applied loss was 0, 1.46, and 3.88 dB, respectively. When the loss is 0 dB, reflection peaks at the PC connectors at 0 and 4.6 km are observed. The reflection peaks at APC connectors does not appear, but the coupling loss at the APC connector at 7.6 km was detected. The sinking of the noise floor after the peak is caused by the intentionally induced loss. We demonstrate with this experiment that detection of reflective surfaces and estimation of fiber transmission loss can be achieved at the same time in AOM-free OCDR systems with degraded spatial resolution.

In conclusion, we have successfully estimated the fiber transmission loss on the basis of the observation of the Rayleigh scattering signal in the AOM-free OCDR and demonstrated that the Rayleigh scattering measurement can be enhanced with degraded spatial resolution. We also experimentally showed the capability of the AOM-free OCDR to simultaneously detect the reflective surfaces in the FUT and estimate the fiber transmission loss. These results indicate that the AOM-free OCDR, similar to the well-established techniques of OTDR and OFDR, can perform detection of both Fresnel reflection and Rayleigh scattering simultaneously during distributed reflectivity measurement, laying a solid foundation for the further application and commercialization of the AOM-free OCDR in the future.

## Acknowledgments

This work was partially supported by the Japan Society for the Promotion of Science (JSPS) KAKENHI (Grant Nos. 21H04555, 22K14272, and 20J22160) and the research grants from the Takahashi Industrial and Economic Research Foundation, the Yazaki Memorial Foundation for Science and Technology, and the Konica Minolta Science and Technology Foundation.

# Figure Captions

**Fig. 1.** Experimental setup of the standard OCDR. AOM: acousto-optic modulator, ESA: electrical spectrum analyzer, FUT: fiber under test, LD: laser diode, PD: photodetector.

**Fig. 2.** Schematic illustration of the difference in observed reflection spectra using **(a)** standard OCDR and **(b)** AOM-free OCDR.

**Fig. 3.** Experimental setup of the AOM-free OCDR. AC: alternating current, DC: direct current, EDFA: erbium-doped fiber amplifier, FG: function generator.

**Fig. 4. (a)** Structure of the ~1 km FUT. **(b)** Reflected power distributions near the FUT end at varied modulation amplitudes (normalized and Lorentzian fitted). The dotted line indicates a 3 dB decrease. **(c)** Dependence of the estimated spatial resolution on the modulation amplitude. **(d)** Reflected power distributions measured along the entire FUT at varied spatial resolutions (normalized). **(e)** Dependence of the Rayleigh scattering power on the spatial resolution.

**Fig. 5. (a)** Structure of the ~10 km FUT. **(b)** Reflected power distribution measured along the entire ~10 km FUT and its linear fitting.

**Fig. 6. (a)** Structure of the FUT consisting of the 0.50 m, 4.6 km, 0.50 m, 3.0 km, and 2.0 km fibers. **(b)** Reflected power distributions along the entire FUT measured with changed loss at 4.6 km.

# Figures

Fig. 1.

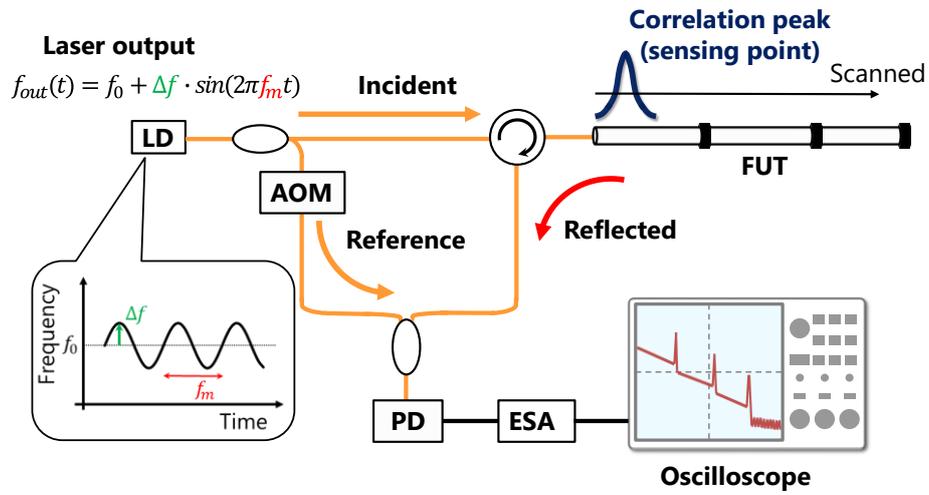

Fig. 2.

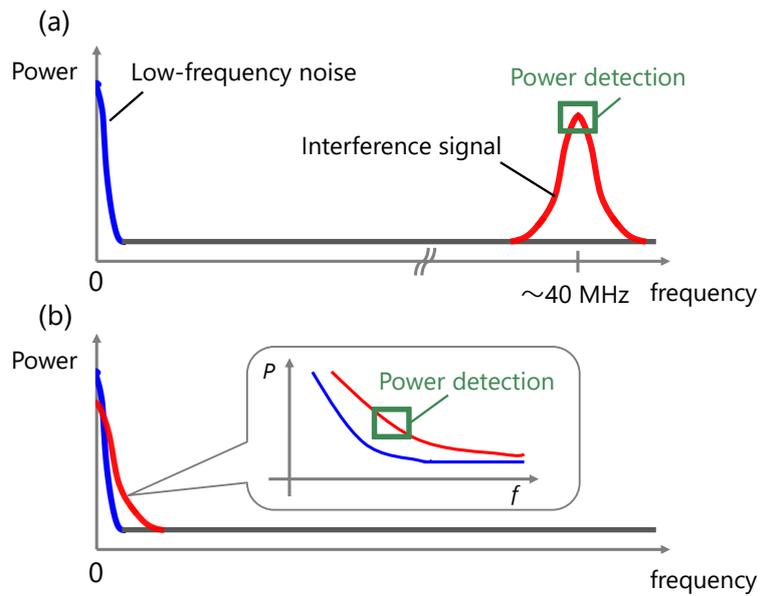

Fig. 3.

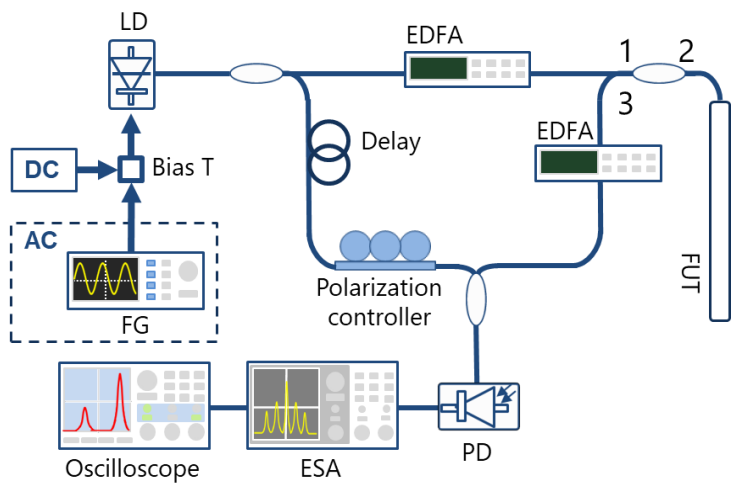

Fig. 4.

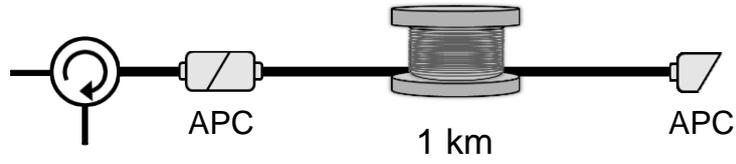

(a)

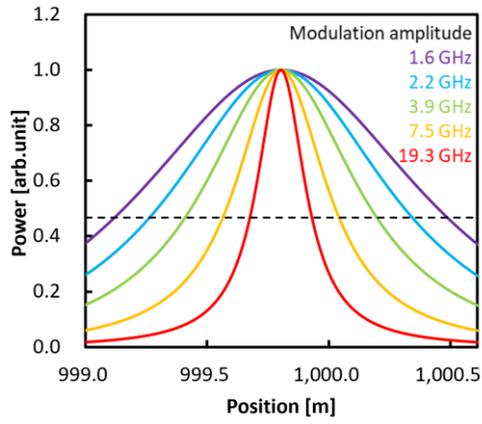

(b)

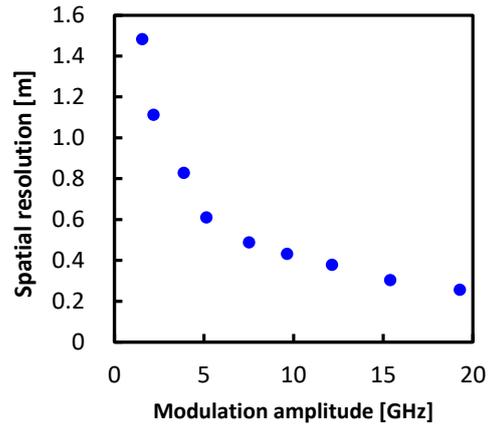

(c)

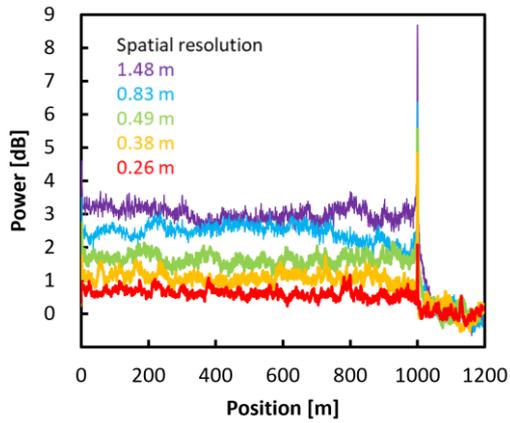

(d)

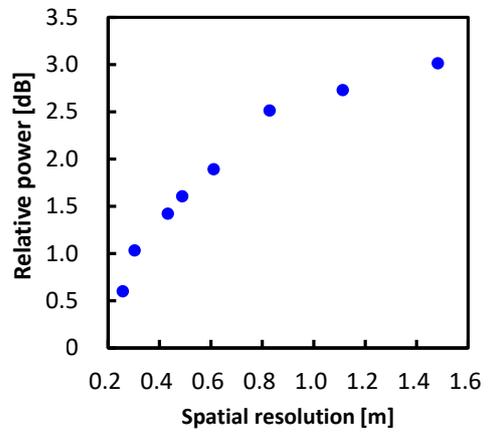

(e)

Fig. 5.

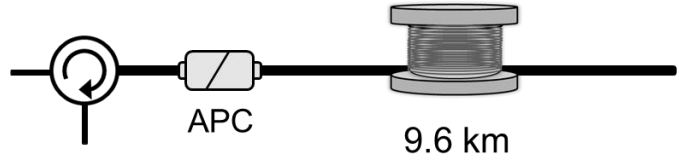

(a)

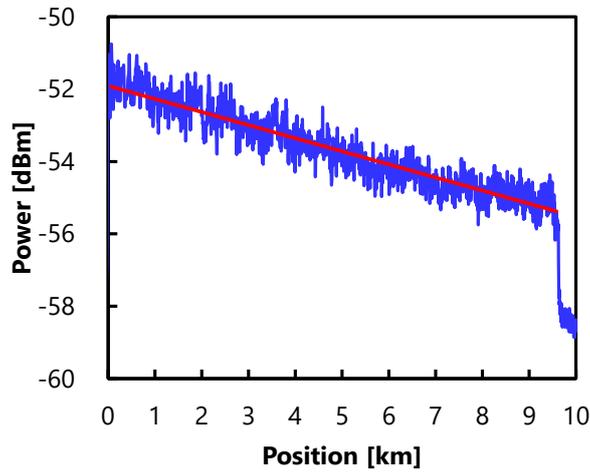

(b)

Fig. 6.

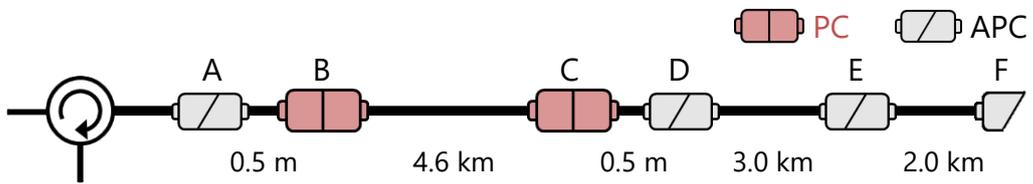

(a)

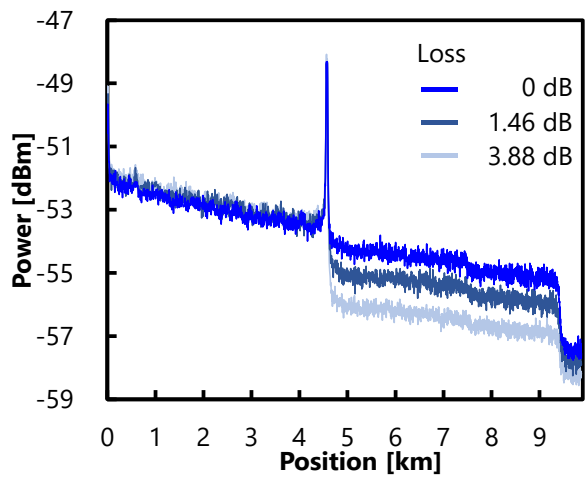

(b)